\documentclass[conference]{IEEEtran}
\IEEEoverridecommandlockouts
\usepackage{cite}
\usepackage{amsmath,amssymb,amsfonts}
\usepackage{algorithmic}
\usepackage{graphicx}
\usepackage{textcomp,epstopdf}
\usepackage{xcolor}
\usepackage{hyperref}
\usepackage{graphicx}
\usepackage{subfigure,color}
\usepackage{changes,setspace}
\usepackage{xcolor}
\graphicspath{ {./images/} }
\newtheorem{lemma}{Lemma}
\setaddedmarkup{\textcolor{blue}{#1}}
\newcommand{\an}{\{a_{n}\}}
\newcommand{\w}{\boldsymbol{\omega}}
\newcommand{\h}{\boldsymbol{h}}
\newcommand{\e}{\boldsymbol{e}}

\newcommand{\Q}{\boldsymbol{Q}}

\newcommand{\F}{\mathcal{F}}

\begin{document}
\title{ Robust Movable-Antenna Position Optimization with Imperfect CSI for MISO Systems}
\author{Haifeng Ma, Weidong Mei, \IEEEmembership{Member, IEEE}, Xin Wei, Boyu Ning, \IEEEmembership{Member, IEEE}, \\and Zhi Chen, \IEEEmembership{Senior Member, IEEE}
\thanks{The authors are with the National Key Laboratory of Wireless Communications, University of Electronic Science and Technology of China, Chengdu 611731, China (e-mails: hfma@std.uestc.edu.cn, wmei@uestc.edu.cn, xinwei@std.uestc.edu.cn, boydning@outlook.com, chenzhi@uestc.edu.cn). This work was supported in part by the National Key Research and Development Program of China under Grant 2024YFB2907900.}}
\maketitle

\begin{abstract}
Movable antenna (MA) technology has emerged as a promising solution for reconfiguring wireless channel conditions through local antenna movement within confined regions. Unlike previous works assuming perfect channel state information (CSI), this letter addresses the robust MA position optimization problem under imperfect CSI conditions for a multiple-input single-output (MISO) MA system. Specifically, we consider two types of CSI errors: norm-bounded and randomly distributed errors, aiming to maximize the worst-case and non-outage received signal power, respectively. For norm-bounded CSI errors, we derive the worst-case received signal power in closed-form. For randomly distributed CSI errors, due to the intractability of the probabilistic constraints, we apply the Bernstein-type inequality to obtain a closed-form lower bound for the non-outage received signal power. Based on these results, we show the optimality of the maximum-ratio transmission for imperfect CSI in both scenarios and employ a graph-based algorithm to obtain the optimal MA positions. Numerical results show that our proposed scheme can even outperform other benchmark schemes implemented under perfect CSI conditions.
\end{abstract}
\begin{IEEEkeywords}
Movable antennas, channel state information (CSI), CSI error, robust antenna position optimization, Bernstein-type inequality.\vspace{-9pt}
\end{IEEEkeywords}

\section{Introduction}
Over the past decades, large-scale multiple-input multiple-output (MIMO) technology has become a pivotal advancement in wireless communications\cite{wang2019overview}. However, the high radio-frequency (RF) cost associated with large antenna arrays poses a significant challenge in practice, as it significantly increases system complexity and operational expenses. Additionally, these antennas cannot fully capitalize on the spatial degrees of freedom (DoF) due to their fixed positions, which may result in suboptimal communication performance.

To tackle this challenge, movable antenna (MA) has been deemed as a promising solution, by allowing multiple antennas to move continuously within a confined region\cite{zhu2024movable,zhu2025tutorial}. By properly optimizing their positions, MAs can significantly enhance the performance of conventional fixed-position antennas (FPAs) while maintaining lower RF costs under various system setups, e.g., MIMO or multi-user systems\cite{ma2024mimo,yang2024flexible}, cognitive radio\cite{wei2024joint}, coordinated multi-point (CoMP) networks\cite{hu2024movable}, integrated sensing and communications\cite{jiang2025movable}, among others. In addition to antenna position optimization, the authors in \cite{zheng2025rotatable} and \cite{shao20256D} also leveraged the rotation of antenna arrays for further performance enhancement. In the context of MA position optimization, prior research has predominantly employed gradient-based algorithms to optimize MA positions assuming the field-response channel model\cite{zhu2024modeling}. In contrast, the authors in \cite{mei2024movable} discretized the movable region into multiple sampling points and proposed a graph-based algorithm to select an optimal subset of these points as the MA positions.

\begin{figure}
    \centering
    \subfigure{\includegraphics[scale=0.4]{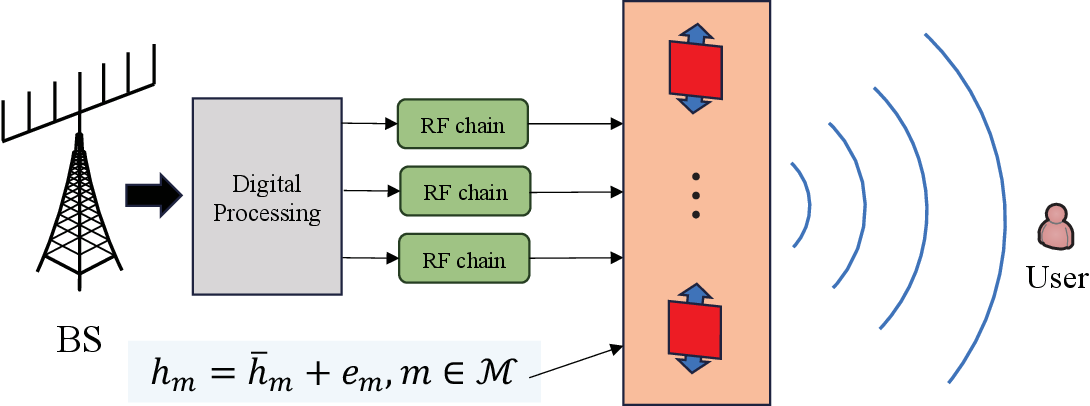}}
    \caption{MA MISO system with imperfect CSI}
    \label{SysMod}
    \vspace{-15pt}
\end{figure}
However, all of the above works assumed perfect channel state information (CSI) at any position within the transmit/receive region to optimize the MA positions. Due to various practical factors, e.g., channel aging and noise effects, it is difficult to acquire perfect CSI in practice. To capture the effects of CSI imperfection on MA systems, we focus on the robust MA position optimization in this letter by considering two types of CSI errors, namely, norm-bounded and randomly distributed errors. The former type assumes that the norm of the CSI error at any position is upper-bounded by a prescribed value, while the latter type models the CSI errors as random variables. Hence, the MA positions are optimized to ensure the worst-case and non-outage performance of a given utility for the first and second types of CSI errors, respectively. Note that although the robust optimization for these two types of CSI errors has been extensively studied in the literature (see e.g., \cite{zhou2020framework} and the references therein), there is no existing work studying the robust antenna position optimization for MAs.

Motivated by the above, this letter addresses the robust MA position optimization problem under imperfect CSI conditions for a multiple-input single-output (MISO) MA system, as shown in Fig.\,\ref{SysMod}. Specifically, we aim to maximize the worst-case and non-outage received signal power at the user for norm-bounded and randomly distributed CSI errors, respectively. For norm-bounded CSI errors, we derive the worst-case received signal power in terms of the CSI error in closed-form. For randomly distributed CSI errors, due to the intractability of the probabilistic constraints, we apply the Bernstein-type inequality to obtain a closed-form lower bound for the non-outage received signal power. Compared to other approximation methods, the Bernstein-type inequality can provide a tighter approximation to probabilistic constraints \cite{wang2014outage}. Based on these results, we show the optimality of maximum-ratio transmission (MRT) for imperfect CSI in both scenarios and propose a graph-based algorithm to obtain the optimal MA positions. Numerical results show that our proposed scheme can even outperform other benchmark schemes implemented 	under perfect CSI conditions.

\begingroup
\allowdisplaybreaks
\section{System Model and Problem Formulation}
\subsection{System Model}
As shown in Fig.\,\ref{SysMod}, we consider an MA-aided MISO communication system, where the transmitter is equipped with $N$ MAs, while the receiver is equipped with a single FPA. A linear transmit array of length $L$ is employed at the transmitter, over which the positions of the $N$ MAs can be flexibly adjusted. Note that the results in this letter are also applicable to a planar transmit array. Let $\mathcal{N}=\{1,2,...,N\}$ denote the set of all MAs. We assume that the channel between the transmitter and receiver is slowly-varying, such that the antenna movement delay has a marginal effect on communication performance. Alternatively, more advanced electronically driven methods can be applied to achieve equivalent antenna movement with negligible delay\cite{zhu2025tutorial}.

For the ease of implementing antenna movement, we uniformly sample the transmit region into $M(M \gg N)$ discrete positions, with an equal spacing between any two adjacent sampling points given by $\delta_{s}=L/M$. Note that this discretization also captures the finite precision of practical antenna movement, e.g., via stepper motors. Hence, the position of the $m$-th sampling point is given by $s_{m}=\frac{mL}{M}, m \in \mathcal{M} = \{1,2,...,M\}$, with $\mathcal{M}$ denoting the set of all sampling points within the transmit region. As such, the position of each MA can be selected as one of the sampling points in $\mathcal{M}$. Let $a_{n}$ denote the index of the selected sampling points for the $n$-th MA. Thus, the position of the $n$-th MA can be determined as $s_{a_{n}}=\frac{a_{n}L}{M}, n \in \mathcal{N}$. To avoid mutual coupling between MAs, we consider a minimum distance between any pair of MAs, denoted as $d_{\min}$. Thus, it must hold that
\begin{equation}\label{min-dis}
    |a_{i}-a_{j}| \ge a_{\min}, \forall i,j \in \mathcal{N}, i \ne j,
\end{equation}
where $a_{\min} = d_{\min}/\delta_{s} \gg 1$. It follows that the MA position optimization is equivalent to selecting $N$ sampling points from $\mathcal{M}$ subject to (\ref{min-dis}).

Denote by $h_m, m \in {\cal M}$ the baseband-equivalent channel from the $m$-th sampling point to the receiver. In practice, a variety of channel estimation techniques can be applied to estimate the channel from any sampling point to the receiver (also known as channel map); see e.g., \cite{zhang2025successive} and \cite{xu2024channel,zhang2024channel} considering continuous and discrete models, respectively. However, due to the receiver noise and quantization errors, there may exist CSI error for each $h_m$. To characterize the CSI error, we define ${\overline{h}}_m$ as the estimated CSI from the $m$-th sampling point to the receiver. Then, we have
\begin{equation}
    h_m = {\overline{h}}_m + e_m, m \in {\cal M},
\end{equation}
where $e_m$ denotes the CSI error at the $m$-th sampling point.

Let $\boldsymbol{h}(\{a_{n}\})=[h_{a_1},h_{a_2},\cdots,h_{a_N}] \in \mathbb{C} ^{N \times 1}$ denote the baseband-equivalent channel from the selected sampling points to the receiver. Accordingly, we define $\boldsymbol{\overline{h}}(\{a_{n}\})=[\overline{h}_{a_1},\overline{h}_{a_2},\cdots,\overline{h}_{a_N}] \in \mathbb{C} ^{N \times 1}$ and $\boldsymbol{e}(\{a_{n}\})=[e_{a_1},e_{a_2},\cdots,e_{a_N}] \in \mathbb{C} ^{N \times 1}$ as its estimate and the estimation error, respectively. As such, we have
\begin{equation}\label{impCSI}
    \boldsymbol{h}(\{a_{n}\}) = \boldsymbol{\overline{h}}(\{a_{n}\})+\boldsymbol{e}(\{a_{n}\}).
\end{equation}

Let $\w \in \mathbb{C} ^{N \times 1}$  denote the transmit beamforming vector, which is subject to $\lVert \w \rVert^2 \le P_{\max}$, with $P_{\max}$ denoting the maximum transmit power. Then, the actual received signal power is expressed as 
\begin{equation}\label{actPw}
    P_{r}(\{a_{n}\})=\left|\w^H\boldsymbol{h}(\{a_{n}\})\right|^{2},
    a_{n} \in \mathcal{N}.
\end{equation}
It is noted from \eqref{actPw} that  $P_{r}(\an)$ is dependent on the CSI error $\boldsymbol{e}(\an)$. 

\subsection{CSI Error Model and Problem Formulation}
In this subsection, we present two generic CSI error models to characterize $e_m, m \in {\cal M}$, which can be applied to any source of CSI errors.

\subsubsection{Norm-Bounded CSI Error}
In the first CSI error model, we consider that the norm of the CSI error is upper-bounded by a prescribed threshold, i.e., $\lVert\boldsymbol{e}(\an)\rVert \le \delta$, where $\delta$ denotes the threshold. Given this model, we aim to maximize the worst-case received signal power in \eqref{actPw}, i.e., $\min\nolimits_{\lVert\boldsymbol{e}\rVert\le \delta}|\w^H\boldsymbol{h}\{a_{n}\}|^{2}$, by jointly optimizing the MA indices $\an$ and the transmit beamforming $\w$. The associated optimization problem is formulated as
    \begin{subequations}\label{pf1}
    \begin{align}
        \max\limits_{\{a_{n}\},\w}& ~\min\limits_{\lVert \boldsymbol{e}\rVert \le \delta}
        \lvert\w^H\boldsymbol{h}(\{a_{n}\})\rvert^{2} \nonumber\\
        {\text{s.t.}}\; &a_{n} \in \mathcal{M}, n \in {\cal N},\label{pf1a}\\
        &\lvert a_{i}-a_{j} \rvert \ge a_{\min},\forall i,j \in {\cal N},i\neq j,\label{pf1b}\\
        &\lVert\w\rVert^{2} \le P_{\max}.\label{pf1c}
    \end{align}
    \end{subequations}
It is observed that compared to our prior work \cite{mei2024movable} assuming perfect CSI, problem (\ref{pf1}) is more involved to solve due to the worst-case formulation and its semi-infiniteness.

\subsubsection{Gaussian Distributed CSI Error}
In the second model, we assume a random CSI error following complex Gaussian distribution, i.e., $\boldsymbol{e} \sim \mathcal{C} \mathcal{N} (0,\sigma^{2}\boldsymbol{I})$, where $\sigma^2$ denotes the variances of the CSI error for each sampling point. 
Due to the random nature of the CSI error, the received signal power in (\ref{actPw}) becomes random as well. As such, we focus on the non-outage performance of the considered system, aiming to maximize the non-outage rate by jointly optimizing the MA indices $\an$ and the transmit beamforming, i.e.,
   \begin{align}
        \max\limits_{\an, \w, R_0}~ &R_0 \nonumber\\
        {\text{s.t.}}\; & {\text{Pr}}_{\boldsymbol{e} \sim {\cal CN}(0,\sigma^2 \boldsymbol{I})}\{\lvert\w^H\boldsymbol{h}(\an)\rvert^{2} 
        \ge R_0 \} \!\ge\! 1-\rho, \nonumber\\
        &{\text{\eqref{pf1a}-\eqref{pf1c}}},\label{pf2}
   \end{align}
where $R_0$ denotes the non-outage received signal power in the presence of the random CSI error, and $\rho$ denotes the prescribed outage probability.

Note that problem (\ref{pf2}) is even more challenging to be optimally solved compared to problem (\ref{pf1}), as there is no closed-form expression of the outage probability therein. In the next section, we will develop efficient algorithms to simplify and approximate problems (\ref{pf1}) and (\ref{pf2}), respectively, such that they become more tractable to tackle.

\section{Proposed Solution to Problem (\ref{pf1})}\label{worstcase}
In this section, we first focus on solving problem \eqref{pf1} by equivalently recasting it as a finite and more tractable form. Specifically, based on \eqref{impCSI}, the objective function of problem \eqref{pf1} can be rewritten as
\begin{equation}\label{3.1}
    \min\limits_{\lVert\boldsymbol{e}\rVert \le \delta}\lvert\w^H\boldsymbol{h}\rvert^{2}=
    \min\limits_{\lVert\boldsymbol{e}\rVert \le \delta}\lvert\w^H(\boldsymbol{\overline{h}}+\boldsymbol{e})\rvert^{2},
\end{equation}
where the explicit dependence of $\boldsymbol{h}$ and $\boldsymbol{e}$ on $\{a_n\}$ is omitted for brevity. Next, note that
\begin{equation}\label{3.2}
    |\w^H(\boldsymbol{\overline{h}}+\boldsymbol{e})|^{2}
    \ge (\lvert\w^H\boldsymbol{\overline{h}}\rvert
    -\lvert\w^H\boldsymbol{e}\rvert)^{2},
\end{equation}
where the equality is achieved when $\lambda \w^H\overline{\h} = \w^H \e$. Here, $\lambda$ is an arbitrary real and negative scalar. Since $\lvert\w^H\boldsymbol{e}\rvert \le \delta\lVert \w \rVert$ given $\lVert\boldsymbol{e}\rVert\le \delta$, we have 
\begin{align}
     \min\limits_{\lVert\boldsymbol{e}\rVert\le \delta}
        \lvert\w^H(\boldsymbol{\overline{h}}+\boldsymbol{e})\rvert^{2}
        &\ge\min\limits_{\lVert\boldsymbol{e}\rVert\le \delta}(\lvert\w^H\boldsymbol{\overline{h}}\rvert
    -\lvert\w^H\boldsymbol{e}\rvert)^{2} \label{min_e}\\
        &\!=\!\begin{cases}
            0,\!\!\!&{\text{if}}\;\lvert\w^H\boldsymbol{\overline{h}}\rvert \!\le\! \delta\lVert \w \rVert, \\
        (\lvert\w^H\boldsymbol{\overline{h}}\rvert\!-\!\delta\lVert \w \rVert)^{2},\!\!\!&{\text{if}}\;\lvert\w^H\boldsymbol{\overline{h}}\rvert \!>\!\delta\lVert \w \rVert.
        \end{cases}\nonumber
\end{align}

It follows from \eqref{min_e} that if $\delta$ is sufficiently large, the worst-case received signal power may be overwhelmed by the CSI error and nulled. Based on \eqref{min_e}, we only need to focus on the second case and can transform the objective function of problem (\ref{pf1}) into
\begin{equation}\label{obj}
    \max\limits_{\an \in \mathcal{M},\w}
    (\lvert\w^H\boldsymbol{\overline{h}}\rvert-\delta\lVert \w \rVert)^{2}.
\end{equation}

For \eqref{obj}, it is obvious to see that for any given MA positions, $a_n, n \in \cal N$, and under the condition $\lvert\w^H\boldsymbol{\overline{h}}\rvert >\delta\lVert \w \rVert$, the optimal transmit beamforming is given by MRT, i.e., $\w(\overline{\h}) = \sqrt{P_{\max}}\overline{\h}/\lVert \overline{\h} \rVert$. Note that with this MRT, the equality condition in \eqref{min_e} can be met by setting $\e = -\delta\overline{\h}/\lVert \overline{\h} \rVert$. 

By substituting $\w (\overline{\h})$ into \eqref{obj}, we can obtain the following optimization problem with respect to (w.r.t.) the MA positions only, i.e.,
    \begin{equation}\label{opt_an}
        \max\limits_{\an}\;
        P_{\max}(\lVert\overline{\h}\rVert - \delta)^{2},\quad
        {\text{s.t.}}\;{\text{\eqref{pf1a},\,\eqref{pf1b}}}.
    \end{equation}
Under the condition $\lvert\w^H(\overline{\h}) \boldsymbol{\overline{h}}\rvert >\delta\lVert \w(\overline{\h})\rVert$, i.e., $\lVert \overline{\h}\rVert > \delta$, it is equivalent to maximize the channel norm $\lVert\overline{\h}\rVert$ or its square $\lVert\overline{\h}\rVert^2$ in \eqref{opt_an}. Note that this channel norm maximization problem has been studied in our previous work\cite{mei2024movable}, which can be optimally solved by invoking a graph-based algorithm, with the computational complexity in the order of ${\cal O}(NM^2)$. Specifically, it can be easily seen that maximizing $\lVert\overline{\h}\rVert^2$ is equivalent to solving the following problem,
  \begin{equation}\label{graph}
   \max\limits_{\an}\,\sum\nolimits_{n=1}^{N}{\lvert {\bar h}_{a_n}\rvert^2},\quad
   \text{s.t.}\;{\text{\eqref{pf1a},\,\eqref{pf1b}}}.
  \end{equation}
Next, we can construct a directed weighted graph to equivalently transform \eqref{graph} into a fixed-hop shortest path problem that can be optimally solved in polynomial time using the dynamic programming. The details are omitted due to the page limit, which can be found in \cite{mei2024movable}. Let $a^{\star}_n$ denote the optimal solution to problem \eqref{graph}. Hence, the optimal value of problem \eqref{pf1} is given by $P_{\max}(\lVert\overline{\h}(\{a^{\star}_n\})\rVert - \delta)^{2}$ if $\lVert \overline{\h}(\{a^{\star}_n\})\rVert > \delta$; otherwise, it is zero. It is noted that thanks to the MA position optimization, the channel norm can be enhanced compared to that with FPAs, which also enhances the robustness against the CSI error for a given $\delta$. More specifically, let ${\mathbf{h}}_{\text{FPA}}$ denote the channel from a multi-FPA BS to the user. It follows that if $\lVert\overline{\h}(\{a^{\star}_n\})\rVert - \delta > \lVert {\mathbf{h}}_{\text{FPA}} \rVert$, then using MAs under imperfect CSI condition will always yield a better worst-case performance than FPAs under perfect CSI condition.

\section{Proposed Solution to Problem \eqref{pf2}}
In this section, we focus on solving problem \eqref{pf2}. Due to the difficulty in dealing with the probabilistic constraints in \eqref{pf2}, we develop a more tractable approximation to it by introducing the well-known Bernstein inequalities. 

\subsection{Convex Approximation of Problem (\ref{pf2})}
First, we can express $|\boldsymbol{\omega}^H\boldsymbol{h}|^{2}$ as
    \begin{align}
        |\boldsymbol{\omega}^H\boldsymbol{h}|^{2}
        & =\boldsymbol{h}^H\w
        \w^H\boldsymbol{h}=(\boldsymbol{\overline{h}}+\boldsymbol{e})^H\w{\w}^H
        (\boldsymbol{\overline{h}}+\boldsymbol{e})\nonumber\\
        & =\boldsymbol{e}^H\boldsymbol{W}\boldsymbol{e}
        +2\Re(\boldsymbol{e}^H\boldsymbol{W}\boldsymbol{\overline{h}})
        +\boldsymbol{\overline{h}}^H\boldsymbol{W}\boldsymbol{\overline{h}},\label{4.1}
    \end{align}
where $\boldsymbol{W}=\w\boldsymbol{\omega}^H$. Let $\e = \sigma\boldsymbol{\xi}$, with $\boldsymbol{\xi} \sim {\cal {CN}}(0,\boldsymbol{I})$. Then, \eqref{pf2} is equivalent to \vspace{-6pt}
\begin{subequations}\label{pf2.1}
    \begin{align}
        \max\limits_{\an,\w, R_0}~&R_0 \nonumber\\
        {\text{s.t.}}\;&{\text{Pr}}_{\boldsymbol{\xi} \sim {\cal CN}(0,\boldsymbol{I})}
        \{\sigma^{2}\boldsymbol{\xi}^H\boldsymbol{W}\boldsymbol{\xi}+2\sigma\Re(\boldsymbol{\xi}^H \boldsymbol{W}\boldsymbol{\overline{h}})
        +\nonumber\\
        &\qquad\quad\boldsymbol{\overline{h}}^H\boldsymbol{W}\boldsymbol{\overline{h}} \ge R_0\} \ge 1-\rho, \label{prob}\\
        &\;{\text{\eqref{pf1a}-\eqref{pf1c}}}.\nonumber\vspace{-6pt}
    \end{align}
\end{subequations}

The main challenge in solving problem \eqref{pf2.1} lies in the probabilistic constraint \eqref{prob}. To tackle this challenge, we let $s=\boldsymbol{\overline{h}}^H\boldsymbol{W}\boldsymbol{\overline{h}}-R_0$, $\boldsymbol{r}=\sigma \boldsymbol{W}\boldsymbol{\overline{h}}$, and $\Q = \sigma^{2}\boldsymbol{W}$, recasting this constraint as 
\begin{equation}\label{4.2}
    {\text{Pr}}\{\boldsymbol{\xi}^H\Q\boldsymbol{\xi}
        +2\Re(\boldsymbol{\xi}^H\boldsymbol{r})
        + s \ge 0\} \ge 1-\rho.
\end{equation}

\begin{lemma}\label{bein}
The inequality in \eqref{4.2} is satisfied if the following constraints are met,
\begin{equation}\label{4.3}
    {\text{Tr}}(\Q)-\sqrt{2\ln(1/\rho)(\lVert\Q\rVert^{2}+2\lVert\boldsymbol{r}\rVert^{2})}+s \ge 0.
\end{equation}
\end{lemma}

The Bernstein-type inequality is utilized in the proof of Lemma \ref{bein}, and interested readers can refer to \cite{wang2014outage} for the details. Note that \eqref{4.3} is convex. By replacing \eqref{4.2} with \eqref{4.3}, we can obtain the following approximation of problem \eqref{pf2.1},
\begin{equation}\label{pf2.2}
        \max\limits_{\an,\w, R_0}~R_0,\quad {\text{s.t.}}\;{\text{\eqref{pf1a}-\eqref{pf1c}}, \eqref{4.3}}. 
\end{equation}

\subsection{Proposed Solution to Problem \eqref{pf2.2}}
As $\boldsymbol{r}=\sigma \boldsymbol{W}\boldsymbol{\overline{h}}$ and $\Q = \sigma^{2}\boldsymbol{W}$, we can obtain 
\begin{equation}
    \begin{aligned}
        &{\text{Tr}}(\Q)={\text{Tr}}(\sigma^{2}\w\boldsymbol{\omega}^H)=\sigma^{2}P_{\max}\\
        &\lVert\Q \rVert^{2}=\lVert\sigma^{2}\boldsymbol{W}\rVert_{F}^{2}
        ={\text{Tr}}(\sigma^{4}\boldsymbol{W}\boldsymbol{W}^H)=\sigma^{4}P_{\max}^{2}\\
        &\lVert{\boldsymbol r}\rVert^{2}=
        \rVert\sigma\boldsymbol{\omega\omega}^H\boldsymbol{\overline{h}}\rVert^{2}
        =\sigma^{2}P^2_{\max}\lvert\overline{\h}^H\boldsymbol{\omega}_{0}\rvert^{2},
    \end{aligned}
\end{equation}
where ${\w}_0={\w}/\sqrt{P_{\max}}$.

Hence, the constraints in \eqref{4.3} can be simplified as
\begin{equation}\label{4.4}
    \begin{aligned}
        \sigma^{2}P_{\max}-\sqrt{2\ln{\frac{1}{\rho}}}\cdot
        \sqrt{\sigma^{4}P_{\max}^{2}+2\sigma^{2}P_{\max}^{2}
        \lvert\overline{\h}^H\boldsymbol{\omega}_{0}\rvert^{2}}\\
        +P_{\max}\lvert\overline{\h}^H\boldsymbol{\omega}_{0}\rvert^{2} \ge R_0.
    \end{aligned}
\end{equation}

Let $y = \lvert\overline{\h}^H\boldsymbol{\omega}_{0}\rvert^{2}$, with $0 \le y \le y_{\max} \triangleq \lVert\overline{\h}\rVert^2$. It is not difficult to verify that for any given MA positions $\{a_n\}$, to maximize $R_0$ in problem \eqref{pf2.2}, it is equivalent to maximize the left-hand side of \eqref{4.4}, i.e.,
\begin{equation}\label{4.5}
    \F(y) = \sigma^{2}P_{\max}-\sqrt{2\ln{\frac{1}{\rho}}}\cdot
    \sqrt{\sigma^{4}P_{\max}^{2}+2\sigma^{2}P_{\max}^{2}y}
    + P_{\max}y,
\end{equation}
which is expressed as a function of $y$.

Next, we derive the first-order and second-order derivatives of $\F(y)$, i.e.,
    \begin{align}
        & \mathcal{F}'(y)=\frac{{\rm{d}}\mathcal{F}(y)}{{\rm{d}} y}=
        P_{\max} - \sqrt{2\ln{\frac{1}{\rho}}}
        \frac{\sigma^{2}P_{\max}^{2}}{\sqrt{\sigma^{4}P_{\max}^{2}+2\sigma^{2}P_{\max}^{2}y}},\nonumber
        \\
        &\mathcal{F}''(y)=\frac{{\rm{d}}^{2} \mathcal{F}(y)}{{\rm{d}} y^{2}} = 
        \frac{\sqrt{2\ln{\frac{1}{\rho}}}\sigma^{4}P_{\max}^{4}}
        {(\sigma^{4}P_{\max}^{2}+2\sigma^{2}P_{\max}^{2}y)^{\frac{3}{2}}} \ge 0.
    \end{align}
It can be seen that $\mathcal{F}'(y)$ monotonically increases with $y$ and satisfies
$
   \mathcal{F}'(0)=
        P_{\max}\Big(1-\sqrt{2\ln{\frac{1}{\rho}}}\Big)$ and $\mathcal{F}'(\infty) 
        \rightarrow P_{\max} > 0
$.
Note that if $0 < \rho < e^{-\frac{1}{2}}$, we have $\mathcal{F}'(0)<0$. While if $e^{-\frac{1}{2}} \le \rho \le 1$, we have $\mathcal{F}'(0)\ge 0$. Moreover, by setting $\mathcal{F}'(y)=0$, we can obtain an extreme point of $\mathcal{F}(y)$ as
$y_0 = \frac{\sigma^{2}}{2}\Big(2\ln{\frac{1}{\rho}}-1\Big)$.

It follows from the above that if $e^{-\frac{1}{2}} \le \rho \le 1$, $\mathcal{F}(y)$ monotonically increases with $y$. While if $0 < \rho < e^{-\frac{1}{2}}$, $\mathcal{F}(y)$ first decreases with $y$ within the interval $[0, y_0]$ and then increases with $y$. Both of the above two cases are illustrated in Figs. \,\ref{dfy} and \ref{fy} for $\mathcal{F}'(y)$ and $\mathcal{F}(y)$, respectively. Note that as a small value of $\rho$ is generally required to achieve a low outage probability, the second case usually holds in practice. Then, a large CSI variance, i.e., $\sigma^2$, may increase the length of the interval $[0, y_0]$ and impair the overall performance.
\begin{figure}
    \centering
    \subfigure[$e^{-\frac{1}{2}} \le \rho < 1$]{\includegraphics[scale=0.45]{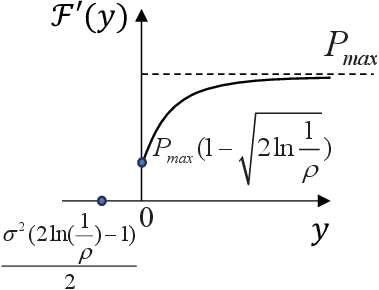}}
    \quad
    \subfigure[$0 < \rho < e^{-\frac{1}{2}}$]{\includegraphics[scale=0.45]{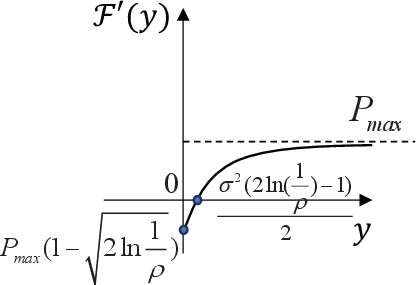}}
    \caption{Illustrations of $\mathcal{F}'(y)$ under different values of $\rho$}\label{dfy}
    \vspace{-9pt}
\end{figure}
\begin{figure}
    \centering
    \subfigure[$e^{-\frac{1}{2}} \le \rho < 1$]{\includegraphics[scale=0.56]{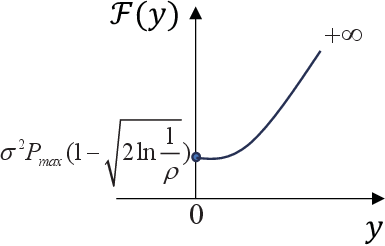}}
    \quad
    \subfigure[$0 < \rho < e^{-\frac{1}{2}}$]{\includegraphics[scale=0.5]{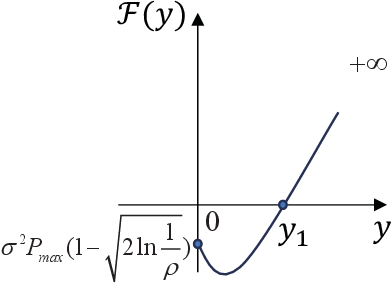}}
    \caption{Illustrations of $\mathcal{F}(y)$ under different values of $\rho$}\label{fy}
    \vspace{-9pt}
\end{figure}

Hence, if $e^{-\frac{1}{2}} \le \rho \le 1$, the maximum value of $\mathcal{F}(y)$ can be attained at $y_{\max}$ with ${\w}_0=\overline{\h}/\lVert\overline{\h}\rVert$. This implies that the MRT should be adopted similarly to the case with norm-bounded CSI error. However, if $0 \le \rho \le e^{-\frac{1}{2}}$, the value of $y$ that maximizes $\mathcal{F}(y)$ depends on the relationship between $y_{\max}$ and the zero point of ${\cal F}(y)$. Let $y_1$ be the zero point of ${\cal F}(y)$ with $\mathcal{F}(y_{1})=0$. Then, if $y_{\max} \le y_1$, the maximum value of $R_0$ should take a negative value and is achieved at $y_2=\arg\max_{y \in \{0,y_{\max}\}}{{\cal F}(y)}$. It should be mentioned that the negative value is due to the approximation introduced by Lemma \ref{bein}, and the actual received signal power should always be non-negative. Nonetheless, the non-outage communication performance may be severely impaired by the CSI error in this case. On the other hand, if $y_{\max} \ge y_1$, then the maximum value of $R_0$ should be achieved at $y_{\max}$ with the MRT. 

In the case of MRT, to maximize $R_0$, the MA positions should be optimized to maximize $y_{\max}$, i.e., $\lVert \overline{\boldsymbol{h}} \rVert^2$. This is the same channel norm maximization problem as problem \eqref{opt_an}, which thus can be optimally solved by solving \eqref{graph} using the graph-based algorithm proposed in our previous work\cite{mei2024movable}. It is noted that similar to the norm-bounded CSI error, the MA position optimization helps enhance the channel norm $\lVert \overline{\boldsymbol{h}} \rVert^2$, as well as the robustness against the CSI error for a given $y_1$. \vspace{-15pt}
\begin{figure*}[!t]
\centering
\subfigure[Worst-case received SNR versus $\delta^{2}$. ]{\includegraphics[width=0.32\textwidth]{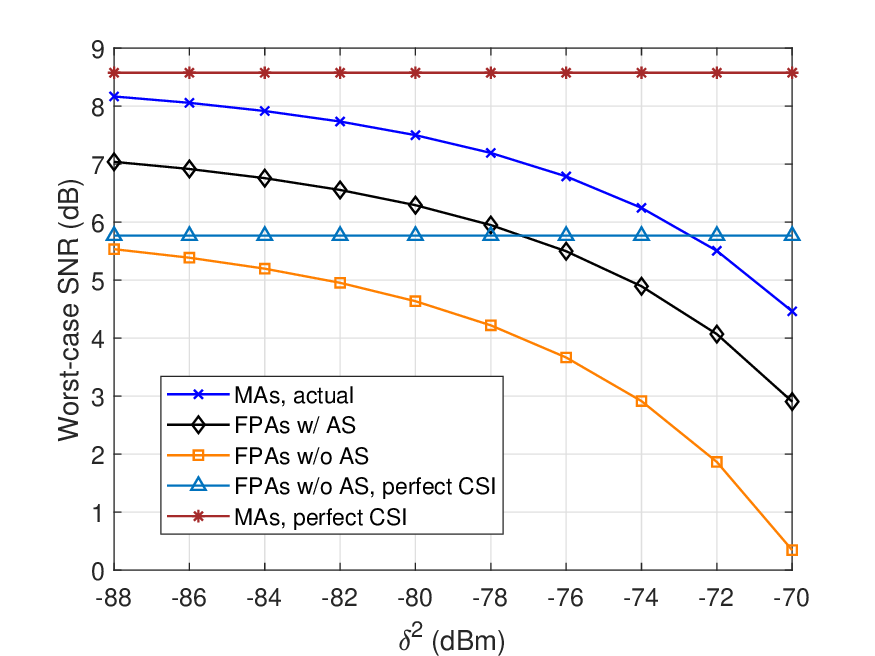}}\!\!
\subfigure[Non-outage received SNR versus $\rho$. ]{\includegraphics[width=0.32\textwidth]{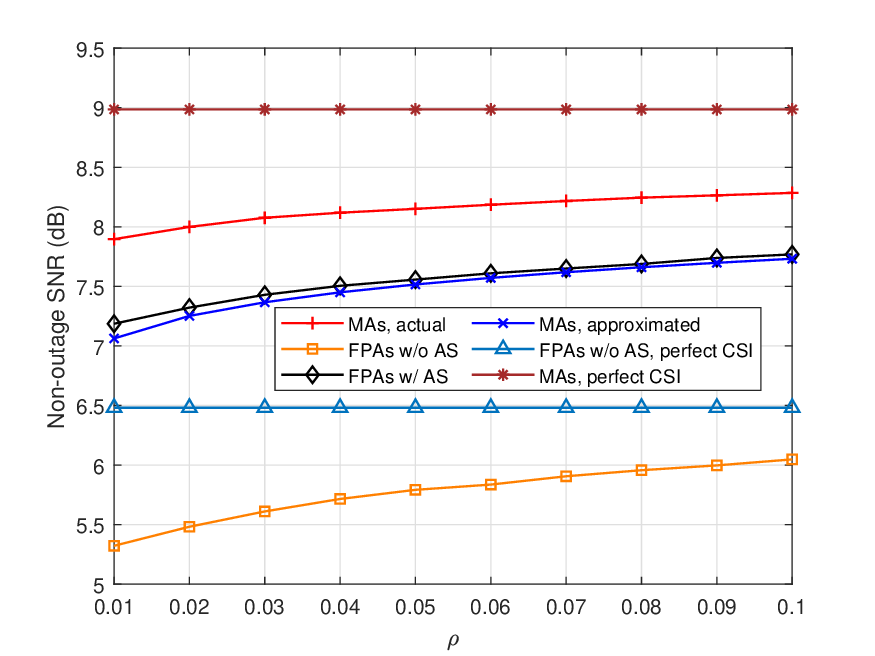}}\!\!
\subfigure[Non-outage received SNR versus $\sigma^2$.]{\includegraphics[width=0.32\textwidth]{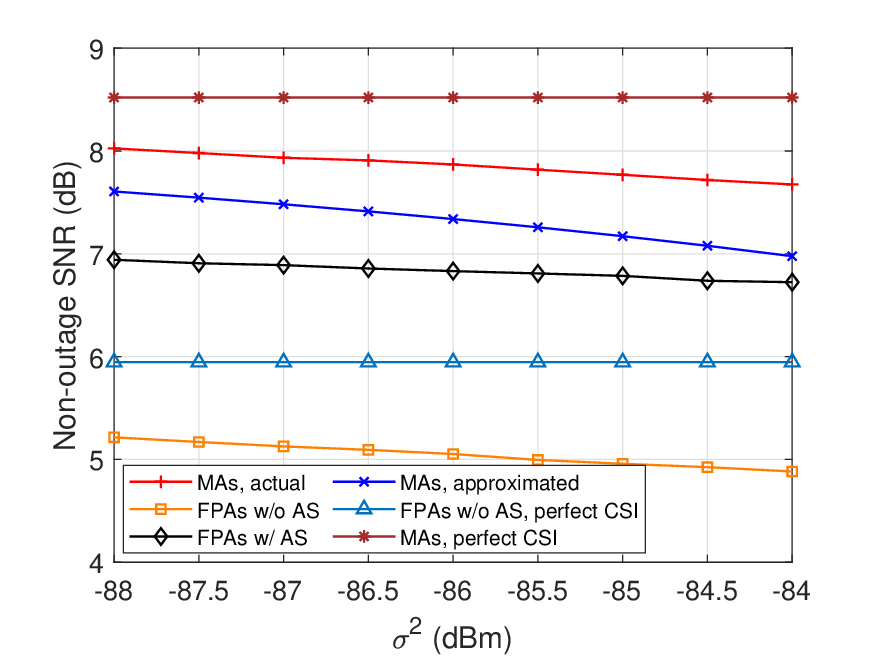}}\!\!
\vspace{-3pt}
\caption{Worst-case and non-outage received SNRs versus different parameters.}\label{simfig}
\vspace{-12pt}
\end{figure*}

\section{Numerical Results}
In this section, we present numerical results to demonstrate the efficacy of our proposed algorithms. Unless otherwise stated, the simulation settings are as follows. The wavelength is $\lambda = 0.06$ meter (m). The number of transmit MAs is $N = 8$, while the length of the linear transmit array is $L = 0.36$ m. The minimum distance between any two MAs is $d_{\min} = \lambda/2$. The distance from the transmitter to the receiver is set to 100 m, and the path-loss exponent is set to 2.8. We consider the field-response channel model as in \cite{zhu2024modeling}, with the number of transmit paths equal to 3 and random power allocations for them. The angle of departure (AoD) from the linear array to each transmit path is assumed to be a uniformly distributed variable over $[0,\pi]$. The transmit signal-to-noise ratio (SNR) is 100 dB.
Moreover, we consider the following two benchmarks, i.e., FPAs without
(w/o) and with (w/) antenna selection (AS). In the FPA w/o AS benchmark, we deploy $N$ FPAs symmetrically to the center of the linear array and separated by the minimum distance $d_{\min}$. In the FPA w/ AS benchmark, $L/d_{\min}=12$ FPAs are deployed along the entire linear array and separated by the minimum distance $d_{\min}$. Among them, $N$ antennas that maximize the received SNR are activated. In addition, we also show the performance of FPAs w/o AS and MAs under perfect CSI. All results are averaged over 100 channel realizations. 

First, we consider the norm-bounded CSI error and plot the worst-case received SNR as derived in Section \ref{worstcase} versus the error bound $\delta^{2}$ in Fig.\,\ref{simfig}(a). It is observed that the worst-case SNRs by all schemes (except FPAs w/o AS and MAs under perfect CSI) decreases with $\delta^{2}$, as expected. Particularly, the worst-case SNR by FPAs w/o AS is observed to be close to 0 dB as $\delta^2=-70$ dBm, implying that the CSI error may significantly impair the communication performance despite the MRT. It is also observed that the proposed algorithm with MAs achieves a considerably better performance than the two FPA benchmarks under imperfect CSI (around 1.5 dB and 4 dB higher for $\delta^2=-70$ dBm). It is also interesting to note that using MAs under imperfect CSI condition can even outperform FPAs w/o AS under perfect CSI condition when $\delta^2 \le -73$ dBm. This observation suggests that MAs exhibit higher robustness against CSI errors compared to FPAs.

Next, we focus on the case of random CSI error and plot in Fig.\,\ref{simfig}(b) the non-outage received SNRs by different schemes versus the prescribed outage probability. In particular, for all considered schemes, we show their actual achieved non-outage SNRs by generating 500 realizations of CSI errors, arranging the resulting SNR values in a descending order and identifying the $100(1-\rho)$-th percentile. In addition, we also show the lower bound of the non-outage SNR achieved by the Bernstein inequality, i.e., the optimal value of problem \eqref{pf2.2}, labelled as ``MAs, approximated''. It is observed that the non-outage SNRs by all considered schemes (except FPAs w/o AS and MAs under perfect CSI) increase with $\rho$. This is expected, as increasing $\rho$ enlarges the feasibility region of \eqref{pf2.1}. It is also observed that the proposed scheme can yield a higher SNR compared to the Bernstein lower bound and other benchmark schemes, implying the effectiveness of our introduced approximation. Furthermore, for $\rho=0.01$, the proposed scheme under imperfect CSI condition can achieve around 1.5 dB higher SNR than FPAs w/o AS under perfect CSI condition. This suggests that MAs can achieve a high robustness against random CSI error as well.

Finally, Fig.\,\ref{simfig}(c) shows the non-outage received SNRs by different schemes versus the CSI error variance, $\sigma^2$. It is observed that the performance of all considered schemes (except FPAs w/o AS and MAs under perfect CSI) decreases with $\sigma^2$, as expected. Nonetheless, compared to Fig.\,\ref{simfig}(a) for norm-bounded CSI error, the random CSI error has less destructive effects on the received SNR. The possible reason is that random CSI error may randomly (either destructively or instructively) deviate from the perfect CSI, while in the worst-case scenario, the error always leads to the most destructive deviation, thus resulting in more significant performance degradation. Moreover, it is observed that the proposed scheme can yield a better performance than all other benchmark schemes, similarly to the observations made from Fig.\,\ref{simfig}(b).\vspace{-4pt}

\section{Conclusion}
This letter focused on the performance optimization of an MISO MA system in the presence of norm-bounded and randomly distributed CSI errors. For the norm-bounded CSI error, we characterized the worst-case received signal power in closed-form. For the randomly distributed CSI error, we applied the Bernstein inequality to yield a tractable lower bound on the non-outage received signal power. It was shown that MRT is optimal for both types of CSI errors in general, based on which a graph-based method was employed to determine the optimal MA positions. Numerical results demonstrated that using MAs can achieve a high robustness against CSI errors and even outperform FPAs implemented under perfect CSI conditions. It would be interesting to extend the results in this letter to more general system setups, e.g., multi-user or MIMO systems.\vspace{-3pt}

\begin{spacing}{0.92}

\end{spacing}
\end{document}